\pdfoutput=1
\documentclass[a4paper,12pt]{article}
\usepackage[english]{babel}
\usepackage{graphics}
\usepackage{graphicx}
\usepackage{parskip}
\DeclareGraphicsExtensions{.pdf,.png,.jpg}
\usepackage{cite}
\title{A Transfer Hamiltonian model for devices based in quantum dot arrays }
\author{S. Illera, J. D. Prades, A. Cirera and A. Cornet\\
\small MIND/IN$^2$UB Departament d'Electr\`onica, Universitat de Barcelona,\\\small C/Mart\'i i Franqu\`es 1, E-08028 Barcelona, Spain}
\date{}
\begin{document}
\maketitle

\hrulefill
\begin{abstract}
We present a model of electron transport through a random distribution of interacting quantum dots embedded in a dielectric matrix to simulate realistic devices. The method underlying the model depends only on fundamental parameters of the system and it is based on the Transfer Hamiltonian approach. A set of non-coherent rate equations can be written and the interaction between the quantum dots and between the quantum dots and the electrodes are introduced by transition rates and capacitive couplings. A realistic modelization of the capacitive couplings, the transmission coefficients, the electron/hole tunneling currents and the density of states of each quantum dot have been taken into account. The effects of the local potential are computed within the self-consistent field regime.\smallskip

While the description of the theoretical framework is kept as general as possible, two specific prototypical devices, an arbitrary array of Qds embedded in a matrix insulator and a transistor device based on Qds, are used to illustrate the kind of unique insight that numerical simulations based on the theory are able to provide.
\end{abstract}
\hrulefill\\
\small Contact author: sillera@el.ub.es

\section{Introduction}
The demand for increasing the integrated density devices has led to the emergency of a whole generation devices based on confined structures. The MOS transistor is the archetype of a confined two-dimensional system \cite{mos}. Nevertheless, the possibility to enhance this confinement by embedding low-dimensional structures in an insulating matrix has opened new way for further downscaling. Compared to the standard bulk technology, the corresponding devices based in these structures have increased the structural and conceptual complexity. These structures (quantum dots, wires or layers) can be used in single-electron devices \cite{Meir}, new memory concepts \cite{tiwari} and photon or electroluminescent devices \cite{lock}. Concerning single electron devices, they are currently conceived to take advantage of tunnel current between quantum states belonging to nanoscale particles \cite{asho, cob}. The single-electron devices based on Qds appear to be potential candidates to improve, to complete or even to replace the current MOS technology with which they may remain compatible. In order to be able to asses the potentials and capabilities of the various novel devices, a realistic theoretical estimate of the specific device performance is thus highly desirable. Within this context, the simulations of such devices must be performed not only to understand but also to predict experimental behaviors. Moreover, from a physical point of view we will learn a lot from these simulations if they are independent on high-level experimental parameters (as tunneling rates, defective interfaces...) and are based on low-level concrete ones (geometrical data, barrier height...). \smallskip

Concerning quantum dots (Qds), they are particularly attractive because they possess discrete energy levels and quantum properties similar to natural atoms or molecules due to the strong confinement in all three directions. This fact affects dramatically the electronic transport properties. Until now, research has mostly concentrated on single Qds and many novel transport phenomena have been discovered, such as the staircaselike current-voltage (I-V) characteristic \cite{Bar}, Coulomb blockade oscillation \cite{Weis}, negative differential capacitance \cite{Wang} and the Kondo effect \cite{vander}.  \smallskip

From experimental point of view, rapid progress in microfabrication technology has made possible coupling quantum dots system with aligned levels \cite{Haug, van, waug}. In fact, the use of the Coulomb blockade phenomenon in systems made up of combinations of tunnel junctions and semiconductor Qds seems to offer promising perspectives, in particular in non-volatile memory applications and also for single-electron transistor \cite{choi}. Moreover, the concept of multi-dot memory using semiconductor nanocrystals embedded in an insulator matrix as floating-gate has already been demonstrated \cite{tiwari2} and the quantization effects have been used in self-aligned double-stacked memory to improve the retention time \cite{Ohba}. \smallskip

From a theoretical perspective, researchers have recently paid much attention to electron transport through several Qds, since multiple Qd provides more Feynman paths for the electron transmission \cite{feyman}. The complexity of structure and physical mechanism as well as the prominent role of dimensional and quantum effects characterizing the operation of these novel Qds devices preclude the use of standard macroscopic bulk semiconductor transport theory. Many authors have studied the electron transport using NEGFF \cite{yeyati,Meir2}, taking into account the potential due to the self-charge. However, up to now nobody has done a computation of transport in an extended arbitrary array of Qds using this framework since this approach is usually unfeasible to implement for large systems. On the other hand, rate equation type models used for lasers or light-emitting diodes often offer a satisfactory description of the charge transport. Moreover, this approach presents a more transparent vision of the electron transport. Thus, this model is easier to tinker with, in order to deal with more complicated nanostructures based on quantum dots. \smallskip

In this work, we present in full a model based in non-coherent rate equations \cite{gur}, which is suitable to study the electron transport in Qd arrays. In a previous work \cite{illera}, a preliminary version of this methodology was presented and used to obtain analytical solutions for electron transport in simple cases. The methodology was also compared with non-equilibrium Green's function calculations, obtaining favorable results \cite{illera2}: despite the simplicity of the model, it provided good results and it was also easily scalable. \smallskip

Now, a complete model to simulate devices based in Qd arrays is presented. The theoretical formalism and the assumptions made in the model are thoroughly described. The interaction between Qds, and between these and the leads has been introduced by transition rates and capacitive couplings. The local potential effects were computed self-consistently. Inelastic and back scattering effects were neglected. Concerning the transition rates, the use of ab initio calculations is shown to be the  best way to fully understand the underlaying tunneling physics in nanostructures. However, if first-principles calculations for single tunnel events were implemented, the huge  effort required would make the simulation time increase in an unacceptable manner. This impractical computational time, forced us to write a compact model with some assumptions and relaxing the expectation of accuracy when treating with few-electron devices operating through quantum features. Specifically, we used the one-dimensional WKB approximation, which neglects spatial variation of the wavefunction over the non-transport directions to describe the tunnel processes. The hole transport was also introduced obtaining new current terms and realistic expressions for the capacity in bipolar conduction. Details about the implementation of the model in the SIMQdot code are also given. This implementation was used to simulate two examples of practical implementations of Qd-based devices: an arbitrary array of Qds operated in conductometric mode and a Qd transistor device with an additional gate electrode.

\section{The model}
\begin{figure}[h!]
\centering{\includegraphics[width=0.5\textwidth]{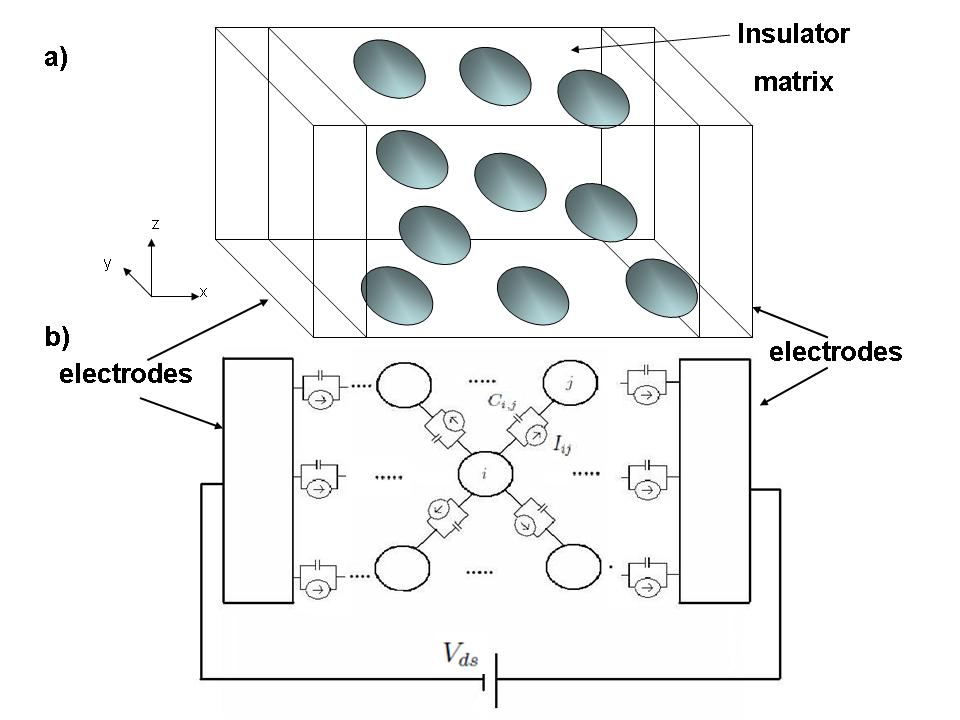}}
\caption{\label{primera}(a) Basic structure and functional elements of the device under study. Qds embedded in an insulator matrix sandwiched between the electrodes. (b) Representation of the system as a network of a multi-tunnel-junctions. }
\end{figure}
As in any device simulation, the ultimate goal of the present approach is to predict the response of a device of one specific architecture (geometry, material...) to a given variation in the external conditions (bias voltage) via the solution of the dynamical equations. First of all, we are going to describe the device architecture and after, we write the underlying equations.

\subsection{The structure}
Fig.~\ref{primera}(a) shows the basic building blocks of our device, which are in essence an insulator layer sandwiched between two metallic electrodes. Inside the insulator layer a random distribution of N quantum dots (Qds) can be inserted. This is the classical structure that is obtained due to the fabrication processes, a superlattice of insulator-semiconductor bilayers. Although single Qd contacted to the leads has been obtained creating a so called single electron transistor (SET) \cite{SET}, the research mainstream is focused on the properties of structures with many Qds to create non-volatile memories \cite{hanafi}, light-emitting devices \cite{jambois} or solar cells devices \cite{solar}. \smallskip

Due to the fabrication processes, the insulator thickness is large enough to avoid direct tunneling between the electrodes (or leads). Therefore, the electron current needs to pass trough the Qds. Thus, a correct modelization of the tunneling processes among the Qds and Qds-electrodes is needed. These tunneling processes can be described by tunnel junctions composed by a capacitance and a current source that depends on the voltage drop. This tunnel junctions network is represented in Fig.~\ref{primera}(b).       

\subsection{Theoretical framework}
A usual method employed to describe the tunneling processes in devices is the tunneling Hamiltonian approach (also called Transfer Hamiltonian approach). This theory, thoroughly studied by many authors \cite{hermann,david,Payne}, treats the tunnelling events as a perturbation. The matrix coefficient $|T_{LR}|^2$ quantifies the probability for a particle to transfer from a state of the left side of the barrier to a state of the right side by a tunnel process.  \smallskip

The tunneling rate is determined using time-dependent perturbation theory by considering the electron from one side of the barrier as initial state and the electron on the other side as a final state. The tunneling rate from the left to the right states (both are considered as a part of a continuum) can be calculated using the Fermi's golden rule \cite{sakurai}
\begin{equation}
\label{uno}
d^2W_{\vec{k}_L \rightarrow \vec{k}_R }=\frac{2 \pi}{\hbar} |T_{LR}|^2 \rho_R (E_R) \rho_L (E_L) \delta(E_R-E_L) dE_R dE_L,
\end{equation}
where $\rho_L$ and $\rho_R$ are the density of states of the left and right side. From this expression, we can see that we only consider ballistic transport. This means that the electron does not suffer energy loss scattering processes when it moves through the barrier. Introducing the energy distribution function in each part of the barrier we can evaluate the total tunneling rate from all occupied states on the left to all unoccupied states on the right \cite{jsee}
\begin{eqnarray}
\label{dos}
\Gamma_{L \rightarrow R} = \int_{-\infty}^{+\infty} \rho_L (E_L) f_L(E_L) \left\{ \int_{-\infty}^{+\infty} \frac{2 \pi}{\hbar} |T_{LR}|^2 [1-f_R(E_R)] \rho_R(E_R) \delta(E_R-E_L) dE_R \right\} dE_L.
\end{eqnarray}   
The opposite tunneling rate can be calculated in a similar way. Thus, the net tunneling current $I=-q [\Gamma_{L \rightarrow R} -\Gamma_{R \rightarrow L}]$ assuming symmetry in the transmission coefficient $T_{LR}=T_{RL}$ \cite{sup1} can be written as
\begin{equation}
\label{tres}
I_{RL}=\frac{4 \pi q}{\hbar} \int_{-\infty}^{+\infty} |T_{RL}(E)|^2 \rho_R(E) \rho_L(E) [f_R(E)-f_L(E)]dE,
\end{equation}
where we introduce a factor 2 to take into account the spin. Therefore, using this approach we can describe the whole system as independent subsystems (the Qds) connected between them by a transmission probability through the dielectric media. Thus, this methodology allows us to write the currents between the different parts of the system. \smallskip

In the above expressions, $f_R$ and $f_L$ are the non-equilibrium energy distribution functions in each side of the barrier. These distributions functions take into account how the energy levels are filled ($f_R$) or emptied ($1-f_R$), as it is expected the electron transport only occurs if the initial state is filled and the final state is empty (see Eq.~(\ref{dos})). However, the distribution function of each part of the system is unknown. Assuming that the distribution functions of the electrodes (left L and right R electrodes) are well described by the equilibrium Fermi Dirac statistics using modified electrochemical potentials, $\mu_L-\mu_R=qV_{ds}$ where $V_{ds}$ is the applied bias voltage, the problem is reduced to find the non-equilibrium distributions functions of each Qd ($n_i$). \smallskip

From the definition of the total charge $N_i$ inside the $i^{th}$Qd can be expressed as: 
\begin{equation}
\label{cuatro}
N_i=\int \rho_i(E) n_i(E) dE.
\end{equation}    
Here, we are going to redefine the notation used to describe the distribution functions. The distribution function for each Qd is $n_i$ while we reserve $f_L$ and $f_R$ for the distribution function of the left and right leads respectively. We can write the evolution charge in time for each Qd as $N_i=\sum_j \int I_{ji}dt$. Where the subscript $j$ takes into account all the elements that are linked to the $i^{th}$Qd. Thus, from Eq.~(\ref{cuatro}) we can write the evolution of charge in time as a function of the total net current flux for each Qd. This set of integro-differential equations have a similar form as a usual rate equations. In the following, the different current terms and the elements that appear in Eq.~(\ref{tres}) are going to be discussed. 

\subsubsection{Electron and holes current terms}
Since the evolution charge in time of each Qd can be written as a function of the net current flux, it is needed to determine all the current contributions. The current contributions can be of two types, the Qds have leads contributions and also neighbors Qd current contributions. These two types of current have the same form in Eq.~(\ref{tres}), but in each case we need to use the correct distribution function. \smallskip

\begin{figure}[h!]
\centering{\includegraphics[width=0.5\textwidth]{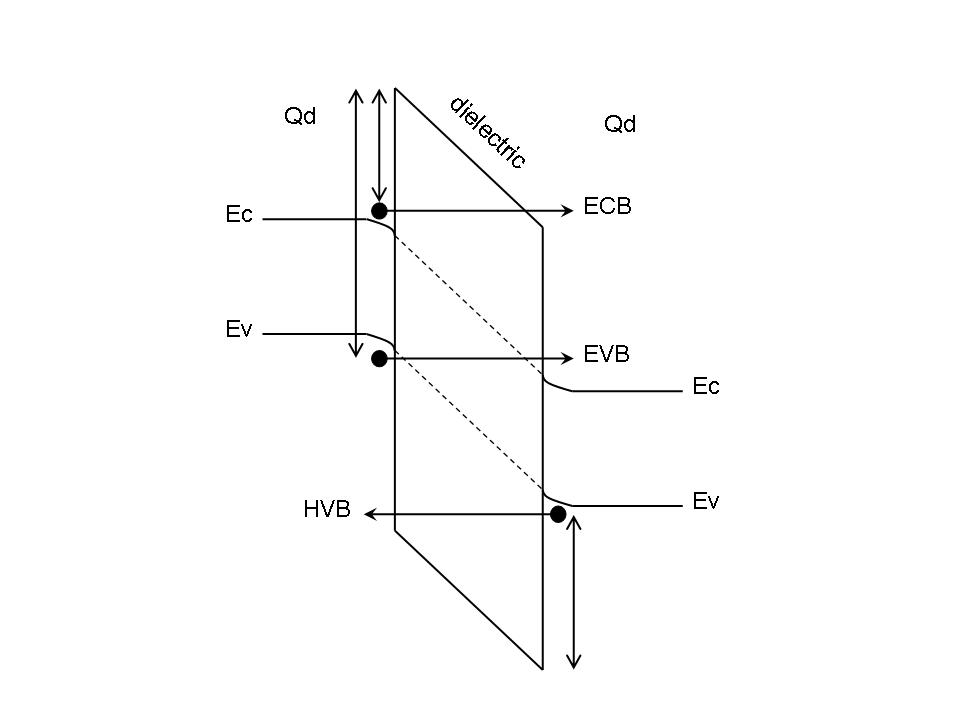}}
\caption{\label{segunda} Schematics of the different tunneling processes. Electron from conduction band to conduction band (ECB), electron from valence band to conduction band (EVB) and tunneling from valence band to valence band (HVB) processes.}
\end{figure}
From the point of view of the nature of these contributions, we have three different processes \cite{lee}. In Fig.~\ref{segunda} we show an scheme of the different tunneling processes. The first term corresponds to electron tunneling from conduction band to conduction band (ECB). The second one is an electron tunneling from valence band to conduction band (EVB). Since the transmission coefficients are symmetric, this process also involves the inverse case, tunneling from conduction band to valence band. The last process is related to the holes: hole tunneling from valence band to valence band (HVB). \smallskip

One important point is how we treat the distinction between electrons and holes. From a physical point of view, the hole conduction can be viewed as electron conduction restricted to the valence band. Therefore, we can consider only electron transport but taking into account the conduction and valence band contributions to the current. Thus, we only need to consider the changing number of electrons in these two bands. Therefore, the time charge evolution equations for each Qd can be written as
\begin{eqnarray}
\label{cinco}
\frac{dN_i}{dt}=&&\underbrace{\int_{-\infty}^{+\infty}|T_{ECB}|^2\rho_L \rho_i^{CB} (f_L-n_i)dE+\int_{-\infty}^{+\infty}|T_{HVB}|^2\rho_L \rho_i^{VB} (f_L-n_i)dE}_{\mbox{Left lead contribution}} \nonumber \\
&&+\underbrace{\int_{-\infty}^{+\infty}|T_{ECB}|^2\rho_R \rho_i^{CB} (f_R-n_i)dE+\int_{-\infty}^{+\infty}|T_{HVB}|^2\rho_R \rho_i^{VB} (f_R-n_i)dE}_{\mbox{Right lead contribution}} \\
{\small{ \begin{array}{c}
\mbox{Neighboring} \\
\mbox{Qds} \\
\mbox{contribution} \\
\end{array}}}
&& \left \{   \begin{array}{c}+\sum_{j,j\neq i}^{N} \left\{ \int_{-\infty}^{+\infty}|T_{ECB}|^2\rho_i^{BC} \rho_j^{BC}(n_j-n_i)dE +\int_{-\infty}^{+\infty} |T_{HVB}|^2\rho_i^{BV} \rho_j^{BV}(n_j-n_i)dE \right\}\\
+\sum_{j,j\neq i}^{N} \left\{ \int_{-\infty}^{+\infty}|T_{EVB}|^2 \rho_i^{BC} \rho_j^{VB}(n_j-n_i)dE +\int_{-\infty}^{+\infty}|T_{EVB}|^2 \rho_j^{BC} \rho_i^{VB}(n_j-n_i)dE \right\} \\
\end{array} \right.
\nonumber
\end{eqnarray} 
where $i=1 \ldots N$ and we take into account all the contributions for an arbitrary $i^{th}$Qd. The first pair of elements are related to the left lead contribution, the electron and the hole contributions. For simplicity, we assume infinite metallic leads therefore we only write the continuum DOS of the leads ($\rho_L$) meanwhile in the Qd we write the DOS in separately terms, conduction ($\rho_i^{BC}$) and valence ($\rho_i^{VB}$) bands. Similar contribution is obtained for the right lead. In these two contributions we use the Fermi Dirac distribution function to describe the leads with $\mu_L-\mu_R=qV$ electrochemical potentials. In each current term, we use the appropriate transmission coefficient. The last two pairs of current terms represent the current from the neighbor Qds. The subscript 'j' runs over all the Qds except the Qd that we are considering. In these terms we take into account the different processes, tunneling from the conduction band (CB) to conduction band (CB) and tunneling from valence band (VB) to valence band (VB). We also need to describe the tunneling that mix the bands (EVB processes) in two ways, from CB to VB and from VB to CB. As it is easy to see, these processes can not occur at the same time but it is important to take both into account.  \smallskip

The set of equations, Eq.~(\ref{cinco}), can be solved for the steady state. Under our assumption that there is no inelastic scattering, the system can be written in a matrix form and solved for each energy step to obtain the non-equilibrium distribution function for each Qd ($n_i$).

\subsubsection{Transmission coefficients}
\begin{figure}[h!]
\centering{\includegraphics[width=0.5\textwidth]{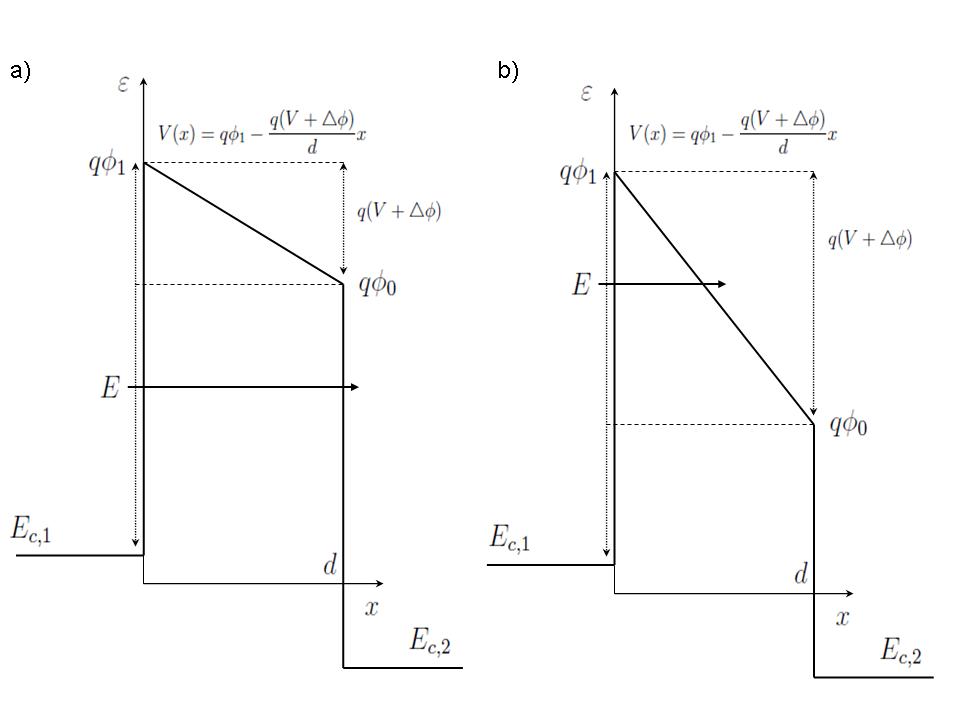}}
\caption{\label{tercera}Energy band diagram for the tunneling processes under polarization. (a) If the incident electron energy (E) is less than the modified energy barrier ($q\phi_0$) we use a direct tunnel expression. (b) If the incident electron energy (E) is greater than the modified energy barrier ($q\phi_0$), we use a Fowler-Nordheim expression. These expressions depend on the incident electron energy but also depend on the polarization voltage. For large polarization voltage a triangular barrier is expected.}
\end{figure}
From Eq.~(\ref{uno}) the transmission coefficient $| T_{LR}|^2$ is defined as the tunnel probability of the electrons pass through the dielectric media. The tunneling probability is a strong function of the parallel component to the junction interface energy $E_{||}$. At each particular total energy E, the DOS with a zero $E_{||}$ component is heavily weighted by the tunneling probability in Eq.~(\ref{tres}). Therefore the DOS only takes into account states with $k_{||}\approx 0$. This approximation may in part be a justification for ignoring the tunneling electron momentum in Eq.~(\ref{dos}) \cite{rus}. \smallskip

Under our assumptions, we use the semiclassical and one-dimension WKB approximation \cite{sakurai} for the transmission coefficient $|T(E)|^2$
\begin{equation}
|T(E)|^2=exp \left\{-\frac{2}{\hbar}\int_{x_1}^{x_2} \sqrt{2 m_{diel}^*(V(x')-E))}dx' \right\},
\end{equation}
where $x_1$ and $x_2$ are the classical turning points, $m_{diel}^*$ is the effective dielectric mass and $V(x)$ is the potential barrier of the dielectric material. This barrier is the difference between the bands of the Qd and the dielectric matrix. The transmission coefficients have been derived  taking into account the effect of the electric field in the interface, $E_{diel}$. The transmission coefficient can be separated in three regions, for incident electrons with  less energy than the modified height of the barrier we use a direct tunnel expression. This expression considers that the electrons see a trapezoidal potential barrier. When the incident electrons have energies between the modified height and the total barrier height we use the Fowler-Nordheim expression in which, the electrons see a triangular potential barrier. Finally, for incident electrons with energy greater than the barrier we do not assume scattering, therefore, we assign  $|T(E)|^2=1$. This last case corresponds to elastic transport trough the conduction band of the dielectric and only occurs for large bias voltages. The transmission coefficient can be written as
\begin{eqnarray}
|T(E)|^2=
\left\{ \begin{array}{ccc}
exp\left\{-4 \frac{\sqrt{2 m_{diel}^*}}{3 \hbar q E_{diel}}((q\phi_1-E)^{3/2}-(q\phi_0-E)^{3/2})\right\}  & \mbox{for}&  q \phi_0 \ge E\\
exp\left\{-4 \frac{\sqrt{2 m_{diel}^*}}{3 \hbar q E_{diel}}(q\phi_1-(E-E_{c,1}))^{3/2}\right\}  & \mbox{for} &q \phi_1 \ge E \ge q \phi_0 \\
 1  & \mbox{for} & E \ge q \phi_1\\
\end{array}
\right. .
\end{eqnarray}
The electric field is defined as $E_{diel}=\frac{E_{c,1}-E_{c,2}+q \triangle \phi}{q d}$, where $q \phi_1$ is the potential barrier height, $q\phi_0$ is the modified potential barrier height, $d$ is the tunneling distance, $E_{c,1 }-E_{c,2}$ is $q$ times the electrostatic potential between the two elements and $q \triangle \phi$ is the work function difference. In Fig.~\ref{tercera} an scheme of the barrier is shown under external polarization. \smallskip

In a similar way the transmission coefficients for the holes can be derived. In that case, the potential barrier is the difference between the valence bands of the Qd and the dielectric. 

\subsubsection{Density of states}
As a first order approximation, we propose a simplified model to represent the discrete energy levels in the Qds. We treat each Qd as a finite spherical potential well. The height of the well is the difference between the conduction band energy level of the dielectric matrix and the one of the material that forms the Qd. \smallskip

Solving the spherical Schrodinger equation inside the well for $l=0$ we obtain the typical binding states \cite{john}. The number of binding states and their energetically position depend on the height of the well $V_0$, the radius $R$ and the electron effective mass $m_{Qd}^*$ and $m_{diel}^*$ (mass inside the Qd and in the dielectric media, respectively). Imposing continuity of the wavefunction and its first derivate in $r=R$ the equation that determine the bounding states are
\begin{equation}
cot x=-\sqrt{\frac{m^*_{diel}}{m^*_{Qd}}}\sqrt{(\frac{\sigma_0}{x})^2-1},
\end{equation}
where $\sigma_0=\sqrt{\frac{2 m_{Qd}^{*}  V_0}{ \hbar^2}(R)^2}$ and $x=\sqrt{\frac{2 m_{Qd}^{*}  }{ \hbar^2}(R)^2 E}$ . This equation can be solved using a Newton-Raphson algorithm that gives us a discrete energy levels $\epsilon_i'$. Since in the Schrodinger equation the zero energy origin is located at the bottom of the well, we need to shift the energy $\epsilon_i'$ in order to have a common Fermi level. Obtaining
\begin{equation}
\rho(E)_i^{CB}=\sum_i^n \delta (E-E_{displ}-\epsilon_i'),
\end{equation}
where $n$ is the number of bounding states in the $i^{th}$ Qd. The value of $E_{displ}$ is half the size of the bulk material gap where we assume the Fermi level is placed. Now, we define $\epsilon_i=\epsilon_i'+E_{displ}$. Similar treatment is done for the valence band to the Qd VB binding states.  \smallskip 

Up to now we treated each Qd as an independent part of the system, but the Qds are coupled between them. This effect is introduced assuming a broadening of the discrete energy levels of the Qds. The standard way to introduce the broadening of the energy levels as a consequence of contacts is to assign a Lorentzian shape to each discrete energy level \cite{Batra}
\begin{equation}
\delta (E-\epsilon) \rightarrow  \frac{\frac{\gamma}{2 \pi}}{(E-\epsilon)^2+(\frac{\gamma}{2 \pi})^2},
\end{equation}
where $\gamma$ is the broadening of the level and it is related with the tunnel probabilities. Therefore, the total density of states (DOS) for each Qd is the total sum of the energy levels taking into account the CB and VB binding states
\begin{equation}
\rho_i=\sum_i^n \frac{\frac{\gamma}{2 \pi}}{(E-\epsilon_i)^2+(\frac{\gamma}{2 \pi})^2}.
\end{equation}
We have used a simplified model in order to describe the DOS structure of the Qds but the proposed approach allows to use more complicated DOS obtained using ab initio models. Therefore, this model is suitable for several materials and an atomistic description of Qds, as we will show in forthcoming papers.

\subsection{Potential profile}
Up to now we have only computed the distribution function of electrons inside each Qd, but the changing in this functions will result into change in the local potential in each Qd ($V_i$). As we can see in Fig.~\ref{primera}(b) each junction is modeled as a current tunnel junction in parallel with a capacity. These capacities represent the electrostatic influence between the different parts of the system. Therefore, each Qd has a local potential due to the applied bias voltage. Since each Qd can be charged we need to solve the Poisson equation 
\begin{equation}
\vec{\nabla} \cdot (\varepsilon_r \vec{\nabla} V_i)=-\frac{q\triangle N_i}{\Omega},
\end{equation}
where $\varepsilon_r$ is the relative permittivity of the dielectric media and $\Omega$ is the Qd volume. $\triangle N_i$ is the change in the number of electrons, calculated respect to the number of electrons $N_0$ initially in the $i^{th}$Qd. The potential energy of each Qd is  $U_i=-qV_i$. The inclusion of the charge term takes into account the carrier interaction inside each Qd. The result of this approach corresponds to the Hartree-term, i.e. first approximation of the carrier-carrier interaction using a mean-field level treatment \cite{suplibro}. \smallskip

The general solution of the potential energy of the $i^{th}$Qd involves the different capacitive coupling between it and its surrounding and its charge increasing \cite{Sup2}, and it can be written as
\begin{equation}
\label{pot}
U_i=\sum_{j\neq i}\frac{C_{ij}}{C_{tot,i}}(-qV_j)+\frac{q^2}{C_{tot,i}} \triangle N_i  ,
\end{equation}
where the subscript $j$ runs over all the components of the system, $C_{ij}$ is the capacitive coupling between the different components and $C_{tot,i}=\sum_{j, j\neq i}C_{ij}$ is the total capacitive coupling of $i^{th}$Qd. The charge energy constant $U_{0i}=q^2/C_{tot,i}$ is the potential increase as a consequence of the injection of one electron into the Qd. Eqs.~(\ref{pot}) are a set of equations (one equation per dot) and the first term of Eq.~(\ref{pot}), the Laplace term $U_i^L$, can be written in a matrix form as:
\begin{eqnarray}
\left( \begin{array}{c} 
U_1^L\\
\vdots\\
U_N^L\\
 \end{array} \right)
=
\left( \begin{array}{ccc} 
1/C_{tot,1}&0&0\\
0&\vdots&0\\
0&\ldots&1/C_{tot,N}\\
\end{array} \right)  \times  [
\left( \begin{array}{c} 
C^{Lead}_1\\
\vdots\\
C^{Lead}_N\\
\end{array} \right) (-qV_{ds}) \nonumber \\
-\left( \begin{array}{cccc} 
0&C_{1,2}&\ldots&C_{1,N}\\
C_{2,1}&0&\ldots&C_{2,N}\\
\vdots&\vdots&\vdots&\vdots\\
C_{N,1}&C_{N,2}&\ldots&0\\
\end{array} \right) 
\left( \begin{array}{c} 
qV_1\\
\vdots\\
qV_N\\
\end{array} \right)  ].
\end{eqnarray}

The first term of the previous equation is the electrostatic influence of the lead in which the bias voltage ($V_{ds}$) is applied meanwhile the second term is the electrostatic coupling with the neighbor Qd. The neighbors capacitive matrix is defined as $N \times N$ symmetric matrix with zero in the diagonal terms. Both terms are multiplied by the inverse of the total Qd capacity. \smallskip

The effects of the local potential on each Qd should be computed in the Qd DOS $\rho_i(E) \rightarrow \rho_i(E-U_i)$ shifting the position of the energy levels. This fact modifies the Qd charge and the currents. In Eq.~(\ref{pot}) we observe that the local potential depends on the increasing charge density but at the same time the charge depends on the DOS that it is modified by the local potential. These considerations impose a selfconsistent solution of Eq.~(\ref{cuatro}) and Eq.~(\ref{pot}). 

\subsubsection{Capacitive elements}
\begin{figure}[h!]
\centering{\includegraphics[width=0.5\textwidth]{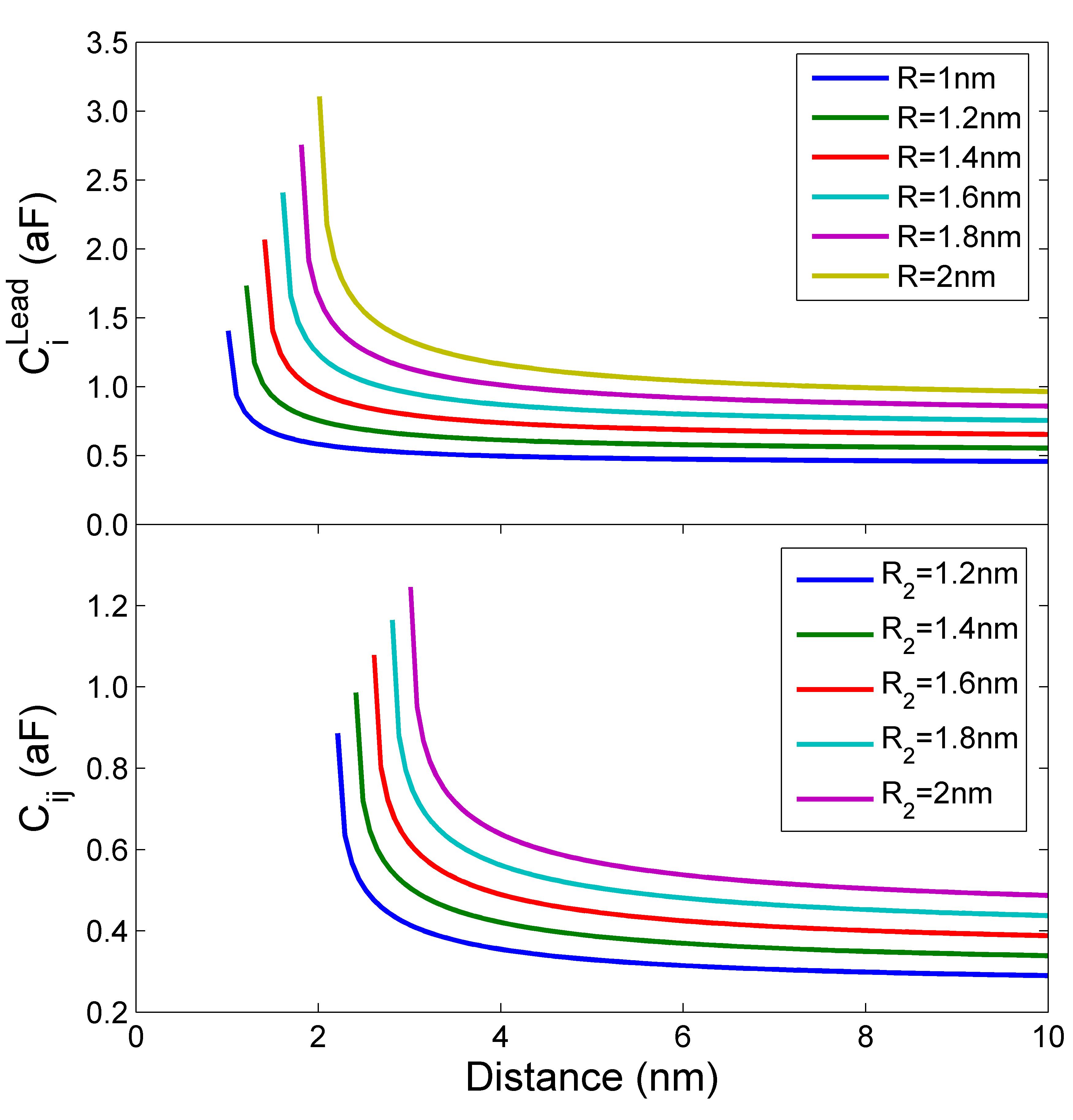}}
\caption{\label{cuarta}(a) Electrode-Qd capacity for different Qd's radii as a function of the distance. (b) Qd-Qd capacity for different $R_2$ radii, the radius of one Qd is hold at $R_1=1nm$. In both plots we use $\varepsilon_r=3.9\varepsilon_0$, where $\varepsilon_0$ is the vacuum permittivity.}
\end{figure}
A realistic modelization of the capacitive coupling between the different parts of the systems \cite{Carreras} is needed, since the electron needs available states in the Qds in order to have transport and the DOS of each Qd depends of the local potential. Therefore, the position of the energy levels with the applied bias voltage plays an important role in the determination of the I-V curve. \smallskip

We use the analytical relationship for a sphere to conducting plane capacitance to model the capacitance between the leads and the Qd, which is
\begin{equation}
\label{quince}
C_{i}^{Lead}=4 \pi \varepsilon_r \sqrt{r^2-R^2}\sum_{n=1}^{\infty}\frac{1}{sinh\,(n\,arccosh(\frac{r}{R}))},
\end{equation}
where $\varepsilon_r$ is the permittivity of the dielectric media, $R$ is the Qd radius and $r$ is the distance between the plane and the center of the Qd.\smallskip

For the case of interdot capacitances ($C_{i,j}$) there is no analytical expression for the capacitance that takes into account Qds of different radii. We use the numerical method of image charges to calculate interdot capacitance between Qds of different sizes. In Fig.~\ref{cuarta} we show the two capacitive terms as a function of the distance and for different Qd radii.

\subsubsection{Self-consistent solution}
\begin{figure}[h!]
\centering{\includegraphics[width=0.5\textwidth]{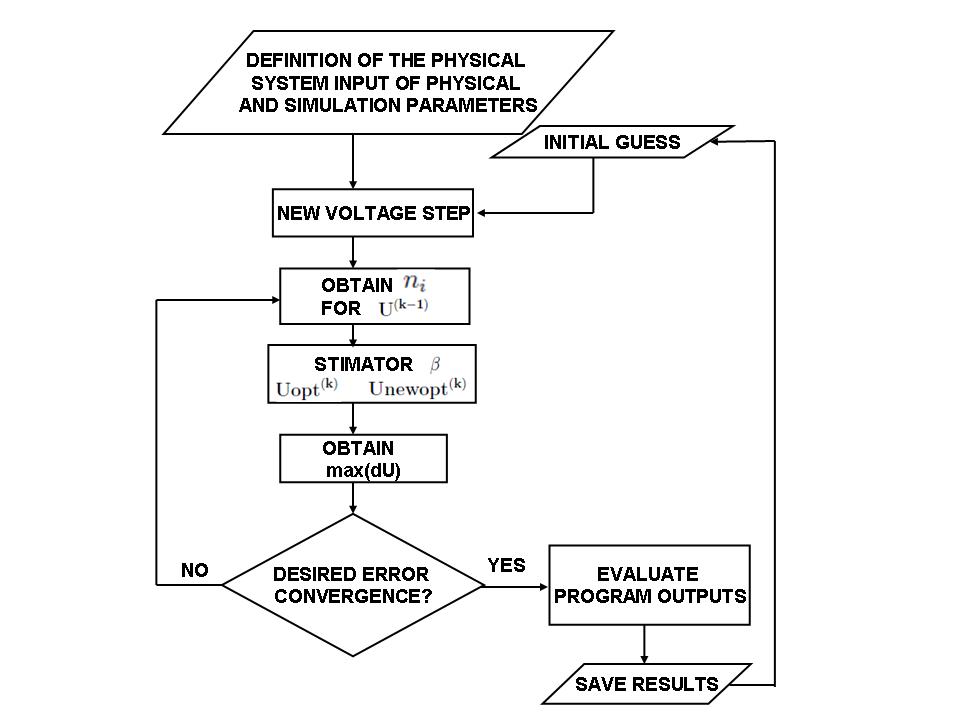}}
\caption{\label{quinta}Scheme flowchart of the implemented model and the Anderson's mixing.}
\end{figure}

Since a simultaneous solution of set of equations Eq.~(\ref{cuatro}) and Eq.~(\ref{pot}) are needed we use numerical methods to achieve the convergence. In Fig.~\ref{quinta} we show the flowchart of the methodology. First of all, a correct description of the device and all the physical parameters are needed. Then, a bias voltage is applied to the electrodes. The Laplace potential ($U^L_i$) is obtained and the transmission coefficients are evaluated. The Qds DOS are shifted and the system of Eq.~(\ref{cinco}) is solved for this local potential. The charge in each Qd is computed ($N_i$) and the solution of the Poisson equation is obtained. An Anderson's mixing \cite{anderson} is used in order to obtain the optimized local potential $U_i$. The Anderson's mixing has been implemented as follows:\\

while dU$>$1*10E-9\\
$\vdots$\\
${\bf{Unew^{(k)}}}={\bf{U0}}\otimes{\bf{(N-N0)}}$\\
$\beta=\frac{(({\bf{Unew^{(k)}-U^{(k-1)}}})\cdot({\bf{Unew^{(k)}-U^{(k-1)}-Unewold^{(k-1)}+Uold^{(k-1)}}}))}{(({\bf{Unew^{(k)}-U^{(k-1)}-Unewold^{(k-1)}+Uold^{(k-1)})}}\cdot{\bf{(Unew^{(k)}-U^{(k-1)}-Unewold^{(k-1)}+Uold^{(k-1)}))}}}$\\
${\bf{Uopt^{(k)}}}=(1-\beta){\bf{U^{(k-1)}}}+\beta{\bf{Uold^{(k-1)}}}$\\
${\bf{Unewopt^{(k)}}}=(1-\beta){\bf{Unew^{(k)}}}+\beta{\bf{Unewold^{(k-1)}}}$\\
${\bf{Uold^{k}}}={\bf{U^{(k-1)}}}$\\
${\bf{U^{(k)}}}={\bf{Uopt^{(k)}}}+0.1({\bf{Unewopt^{(k)}-Uopt^{(k)}}})$\\
${\bf{dU}}=|{\bf{Unew^{(k)}-U^{(k)}}}|$\\
${\bf{Unewold^{(k)}=Unew^{(k)}}}$\\
dU=max({\bf{dU}})\\
$\vdots$\\
end\\

All the bold used variables are vectors except $\beta$ and dU, which are scalars. The "$\cdot$" symbol represents the scalar product meanwhile "$\otimes$" is the element by element multiplication. To summarize the Anderson method, it is based on the search for the best $\beta$ value which minimizes the distance between the two "average" potentials ${\bf{Uopt^{(k)}}}$ and ${\bf{Unewopt^{(k)}}}$. Finally, to obtain the new guess for the next iteration ${\bf{U^{(k)}}}$ a simple mix is used. We want to remark that all the process is done at the same time for the N Qds and finishes when the convergence is achieved in all Qds. In order to ensure this condition, we use the maximum error of {\ttfamily {dU}} as a criteria to stop the loop. \smallskip

Once the convergence has been achieved, the outputs have been obtained for this voltage step. The process repeats until all the bias voltage steps have been done using the previous potential results as the initial guess for the next bias voltage iteration.\smallskip

The methodology has been explained in depth to able the interested reader to creates his own code and reproduce the following results. However, implementation of the code for specific devices is available \cite{codigo}.

\subsubsection{Code implementation}
\begin{figure}[h!]
\centering{\includegraphics[width=0.5\textwidth]{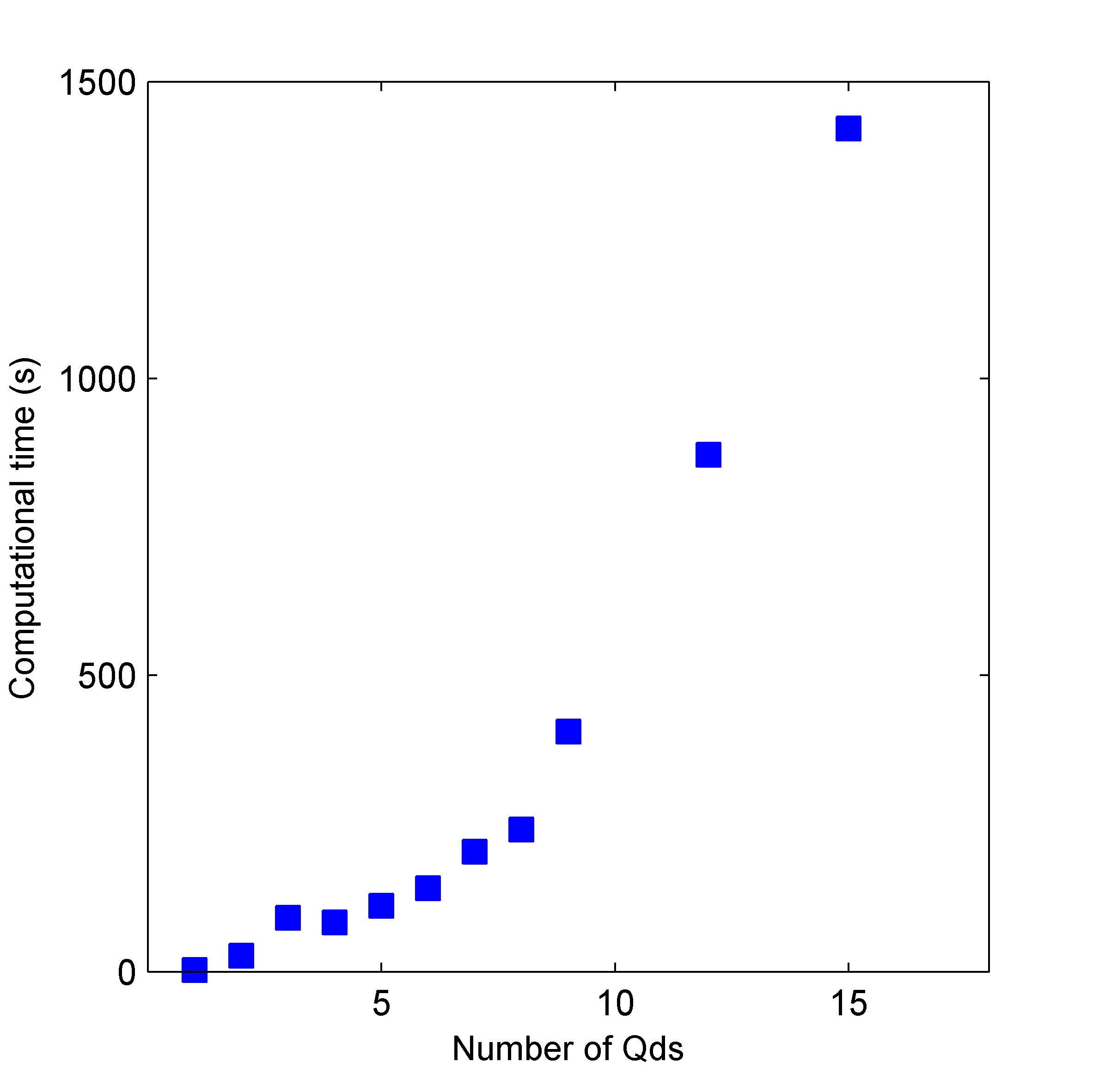}}
\caption{\label{decima} Computational time vs. number of simulated Qds. The time is referred for a single voltage point.}
\end{figure}
The previous formalism has been implemented in MATLAB\copyright{} taking advantage its advantageous matrix-oriented syntax. The general strategy behind the code is presented below. First, the Qds are distributed randomly inside the insulator matrix imposing two conditions: 1) no overlapping between Qds is allowed and 2) the Qds must lie entirely in the insulator. These conditions ensure real and positive capacitive values.\smallskip

In a second stage, a loop is created in order to solve the problem at each voltage point. All the parameters that depend on the applied voltages are calculated (i.e. transmission coefficients) and the set of equations is created. The Eq.~(\ref{cinco}) are solved using the Moore-Penrose pseudoinverse matrix method, since the Eq.~(\ref{cinco}) matrix might not always be directly invertible. The charges are calculated and the Anderson mixing is used. This process is repeated until the desired convergence is achieved.\smallskip

In a final staged part, the outputs (Qd occupancy vs. voltage, current vs. voltage, local potential vs. voltage) are calculated for each voltage point and  saved in a matrix structure.\smallskip

In Fig.~\ref{decima} we show the computational time needed to obtain results for one voltage point as a function of the number of Qds. The computational time grows with the number of Qds but it is still reasonable and allows to simulate large Qd arrays. Moreover, MATLAB\copyright{} allows to execute the code in parallel. Therefore, the simulations can be run in a parallel cluster decreasing further the computational time.

\section{Applications}
In the following, the information provided by the previous formalism will be illustrated studying two different generic devices based on Qds: (1) a random distributed array of silicon (Si) Qds embedded in a silicon dioxide ($SiO_2$) insulator matrix placed among two electrodes, and (2) a similar structure but with a third electrode placed on top of the device, a transistor structure. While the former device embodies all relevant electron transport mechanism, the latter corresponds to a prototypical application of this kind of systems. \smallskip

The two devices are fundamentally similar in several aspects of their working principles. The electron transport takes place from the left electrode (source) to the right electrode (drain) through the Qds. For a physical point of view, the most general condition in order to obtain transport is that the energy levels of the Qds must lie between the electrochemical potentials of the leads ($\mu_L-\mu_R=qV_{ds}$), this condition can be summarized as $\mu_L \ge \epsilon_i \ge \mu_R$. The type of transport will depend on the nature of these energy levels. We have considered all the tunnel current between the $i^{th}$ Qd among its surrounding, but since the transmission coefficient is strongly dependent with the tunneling distance, some processes are more favored than others. Therefore, other transport conditions appears. In order to have transport between the Qds, overlapping of the Qd DOS is necessary. Free states in the arriving Qd are also needed. Thus, the systems plays with the transmission probabilities between the different processes and the available states. Therefore, the total net current will be the sum of the partial tunnel currents among the Qds and the right lead. This current is going to be dependent on the position of the Qds (the tunneling distances) and the alignment of the energy levels (the local potential and the DOS of each Qd).
\begin{table}[h!]
\begin{center}
\begin{tabular}{|c |c| |c| c|}
\hline
$m^*_{ECB}$ ($m_0$) & $0.4$ & $\phi_{1,ECB}$ (eV) & 3.1\\
$m^*_{EVB}$ ($m_0$) & $0.3$ & $\phi_{1,HVB}$ (eV) &-4.5\\
$m^*_{HVB}$ ($m_0$) & $0.32$ & $E_{displ,CB}$ (eV)& 0.6\\
$m^*_{Qd,CB}$ ($m_0$) & $1.08$ &  $E_{displ,VB}$ (eV)& -0.6\\
$m^*_{Qd,VB}$ ($m_0$) & $0.57$ & $\varepsilon_r$ ($\varepsilon_0$) & 3.9\\
\hline
\end{tabular}
\caption{\label{tabla}Parameters used in the simulation in order to describe Si Qds embedded in $SiO_2$ insulator matrix.}
\end{center}
\end{table}\\
In the transistor device, the third electrode (gate) plays an important role. This electrode adds a new term to the local potential (Laplace term) and also introduce a new capacitive coupling between the Qds and the gate. As usually happens for insulated gate-driven devices, we neglect the current among the Qds and the gate. Thus, the gate electrode moves the position of the energy levels of the Qds changing the electrical behavior of the system. \smallskip

The parameters used in the simulations are listed in Table 1. The $m^*_{ECB}$, $m^*_{EVB}$ and $m^*_{HVB}$ are the oxide effective masses for the different tunneling processes, the values are extracted from Lee et al. \cite{lee}. $m^*_{Qd,CB}$ and $m^*_{Qd,VB}$ are the electron and hole Si bulk effective masses used to obtain the binding states in the Qd. We considerer a displacement energy equal to half of Si bulk gap and the potential barriers of the Qd are $\phi_{1,CB}$ for conduction band and  $\phi_{1,VB}$ for valence band. Finally, $\varepsilon_r$ is the the relative permittivity of the $SiO_2$ matrix.       

\subsection{Electron transport through Si Qds array embedded in $SiO2$ matrix }
\begin{figure}[h!]
\centering{\includegraphics[width=0.5\textwidth]{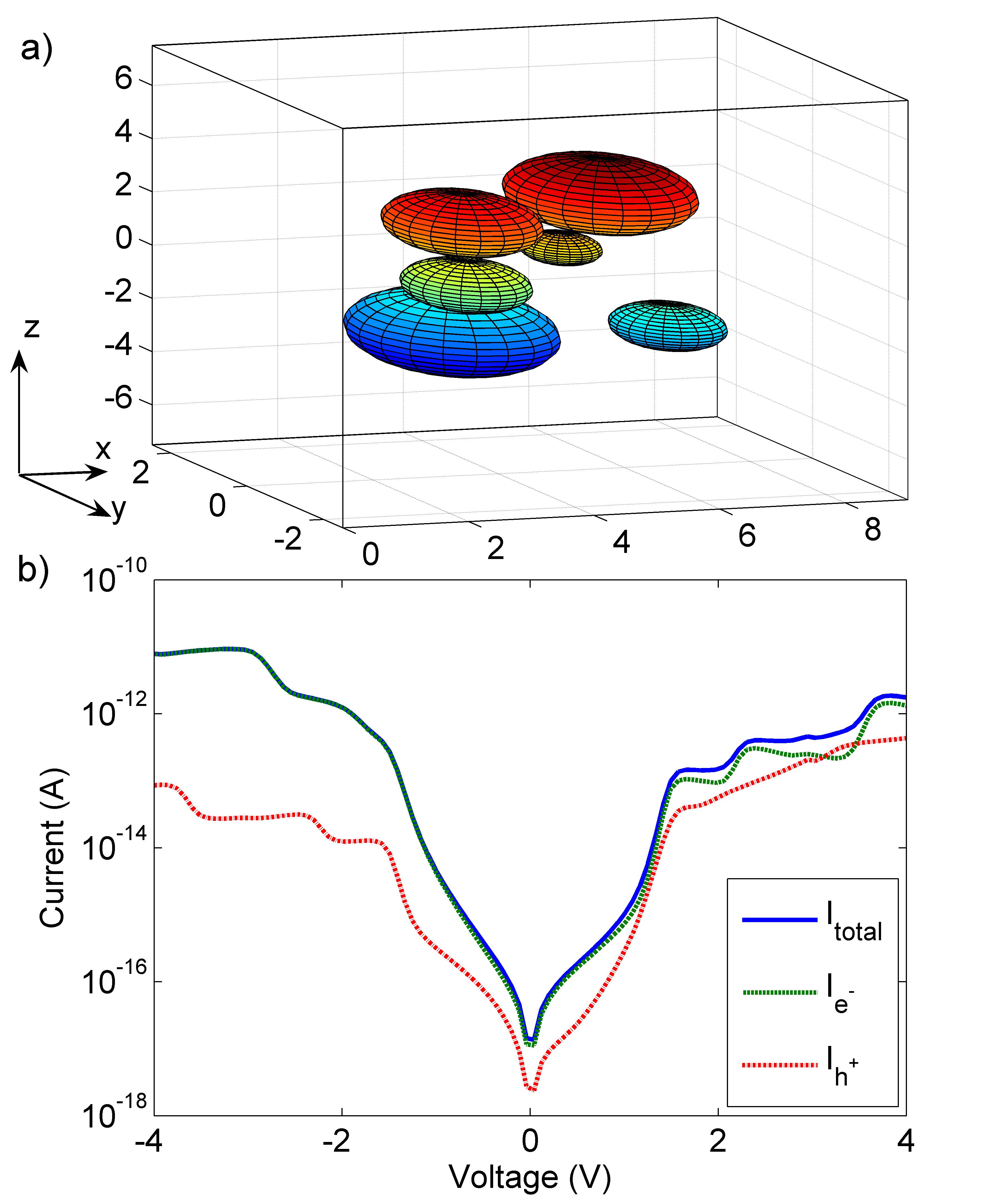}}
\caption{\label{sexta}(a) Structure under consideration. The left and right electrodes are placed in the Y-Z plane at x=0 and Y-Z plane at x=9nm respectively. Six spherical Qds are randomly distributed inside the dielectric matrix with different radii. (b) The total I-V curve (in absolute value) obtained for this system (blue line), the electron current term (green line) and the hole term (red line).}
\end{figure}
In order to deal with devices based in Qds array, first of all we need to study the electron transport in the Qds array. This part is going to be one of the most important building blocks of the device and therefore is important to have a good characterization of it. An scheme of the system under study is shown in Fig.~\ref{sexta}(a) . The system is formed by two electrodes (drain and source), the insulator matrix ($SiO_2$) and $N=6$ Si Qds distributed in the dielectric matrix. The size of the system is 9 nm length, 5 nm width and 10 nm height. The six Qds have been distributed with uniform probability in all the volume and we have used a normal distribution with 1nm mean value and 0.2nm deviation for the Qds radii. \smallskip

The total and electron/hole I-V curve (in absolute value) are presented in Fig.~\ref{sexta}(b). As expected, the hole current term is lesser than the electron current term since the barrier for the hole tunneling is larger. Moreover, the current shows a stepping behavior. This effect is related to the opening of the conduction channels. Negative differential conductance is also obtained as a result of the decreasing overlapping between the conduction energy levels \cite{illera}.   

\subsection{Si Qds transistor}
\begin{figure}[h!]
\centering{\includegraphics[width=0.5\textwidth]{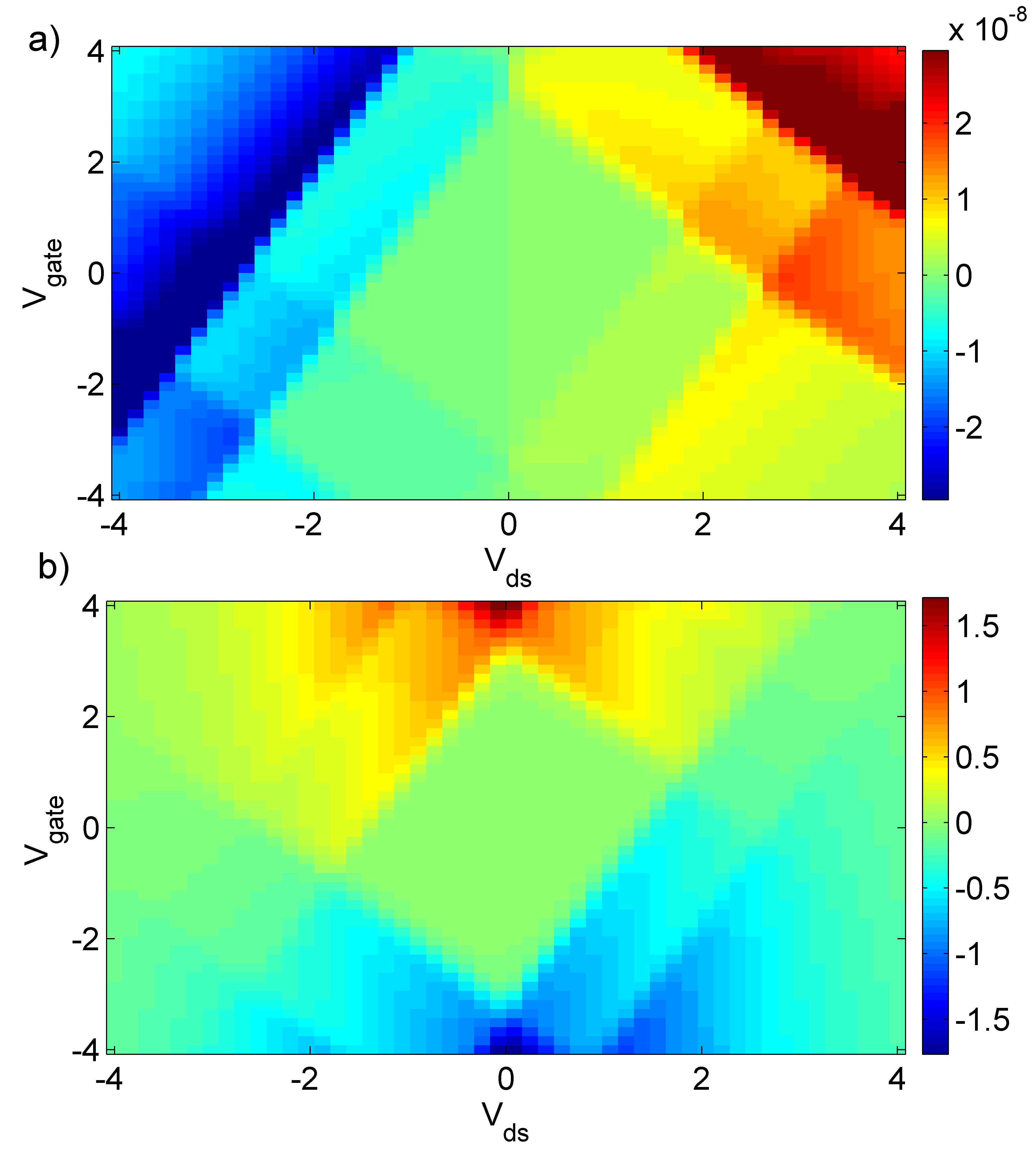}}
\caption{\label{octava}(a) Current map as a function of the applied $V_{ds}$ and $V_{gate}$. Current suppression is obtained until the energy levels are placed between the electrochemical potentials of the drain and source ($\mu_L$ and $\mu_R$). Once a conducting energy level is open the current increase dramatically. (b) Accumulated charge in the Qd map as a function of the applied voltages.}
\end{figure}

The following structure is the typical transistor device. This device is formed by three electrodes (drain, source and gate). This third electrode only changes the local potential of the dots (i.e, moves the DOS) and injects no current. In the Laplace solution we need to add and extra potential term $-C_{gate,i}/C_{tot,i}(-q V_{gate})$ and $C_{tot,i}$ also includes the gate capacity ($C_{gate,i}$). The gate capacity is obtained using Eq.~(\ref{quince}). First, we study a transistor with only one Qd, and after we simulate a transistor with some Qds.
\begin{figure}[h!]
\centering{\includegraphics[width=0.5\textwidth]{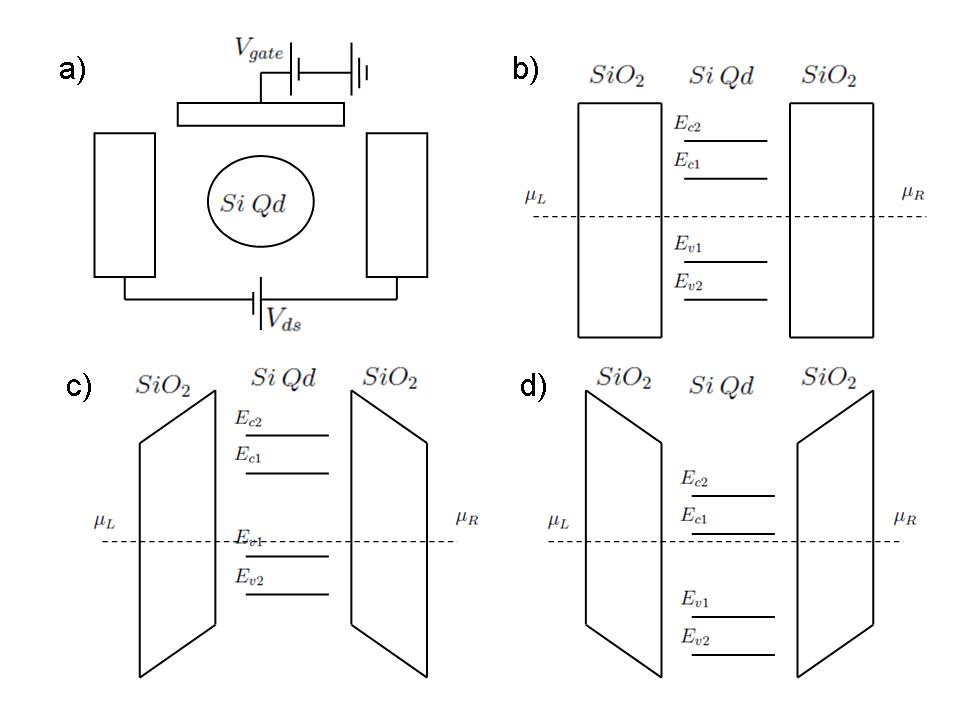}}
\caption{\label{septima}(a) Structure of the system under studding. The scheme shows the three electrodes (source, drain and gate) and the Qd is placed in the middle. The Qd is connected with the source and drain by 1nm and 2nm tunnel junctions respectively. The Qd radius is 1nm. The gate electrode is placed at 7.5nm distance from the center of the Qd. This tunneling distance justifies the assumption that the current between the Qd and the gate is negligible. (b) Band diagram of the structure without applied voltage. The oxide barriers and the conduction and valence energy levels are shown. The equilibrium fermi level of the system is also represented. (c) Energy band scheme of the system under a negative gate polarization. The energy levels are moved upward and the oxide barriers are changed. (d) Energy band scheme of the system under a positive gate polarization.}
\end{figure}\smallskip\\
The scheme of the system is shown in Fig.~\ref{septima}(a). The current is represented as a function of the applied bias voltage (drain-source voltage, $V_{ds}$) and the gate voltage ($V_{gate}$). The stability diagram of the transistor (current curve) and the accumulated charge in the Qd are represented in Fig.~\ref{octava}(a) and Fig.~\ref{octava}(b) respectively. In order to have transport, the energy levels of the Qd must lie between the electrochemical potential of the source and drain lead. In this device, this condition can be achieved as a combination of the applied bias voltage and the gate voltage. \smallskip

For small $V_{ds} \approx 0$ the current is blocked until the first energy level is placed between the electrochemical potentials as a result of the applied $V_{gate}$. For the $V_{gate} \approx 0$ case, the explanation is similar but in this case only $V_{ds}$ contributes to the local potential. Therefore, the current is blocked until the previous condition is achieved. This effect corresponds to the central diamond that it can be seen in the upper figure. \smallskip

The shape of the current diamonds is a result of the relationship between the different capacity values and the position of the energy levels. Once an energy channel is open, the current increase dramatically as a result of the transmission coefficient that dominates the tunneling current. The obtained results are consistent with the results presented in \cite{kastner, Kouwen}. 
We also show the accumulated charge in the Qd. The Qd remains uncharged in a large region since the energy value of the conduction and valence energy levels are similar. Therefore, the Qd begins to charge when the electron transport is the dominant processes. If the hole conduction is preferred the Qd losses its initial charge. This energy level symmetry is broken by the transmission coefficient at high $V_{ds}$. Applying $V_{gate}$ the charge in the Qd is also changed. The physical process is different since $V_{gate}$ moves upward/downward the energy levels across the electrochemical potentials of the leads and the Qd losses/gains charge \cite{kastner}. \smallskip\\
\begin{figure}[h!]
\centering{\includegraphics[width=0.5\textwidth]{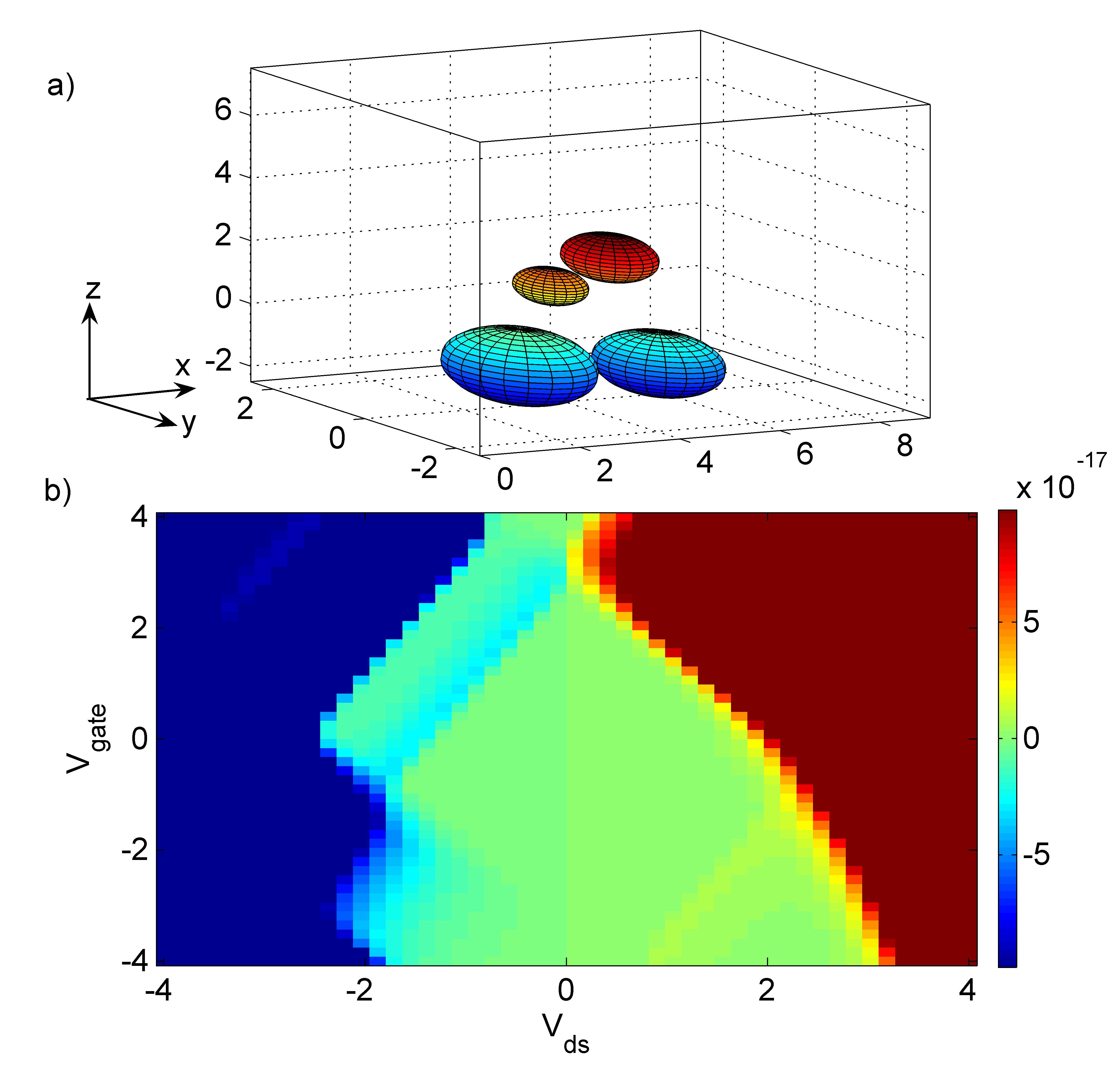}}
\caption{\label{novena}(a) Scheme of the transistor device. The left and right electrodes are placed in the Y-Z plane at x=0nm and Y-Z plane at x=9nm respectively. The gate electrode are placed in the X-Y plane at z=7.5nm. The system is formed by 4 Si Qds randomly generated inside the $SiO_2$ insulator matrix. (b) Current suppression is obtained until the energy levels are placed between the electrochemical potentials of the drain and source ($\mu_L$ and $\mu_R$). Once a conducting energy level is open the current increase dramatically.}
\end{figure}
Once the transistor structure for one Qd has been implemented, the extension of the previous formalism is straightforward to describe the electron transport in an array of Qds. Since there are many Qds the electron has different pathways and the overlapping between the DOS of the Qds appears as a crucial point. Moreover, the gate capacitive coupling depends of the distance between the Qd and the gate and the gate influence in the local potential is not the same in all the Qds. Therefore, the evolution of the energy levels with the applied gate voltage varies opening/closing different conduction channels. The scheme of the simulated system and the corresponding current maps are represented in Fig.~\ref{novena}(a)(b), respectively. 

\section*{Conclusions}
The high efficiency concepts of the next generation of Qds based devices pose new requirements on models for the theoretical description of their transport properties. An intuitive theoretical framework suitable for this purpose is available in the non coherent rate equations. This approach provides a simple and transparent method to describe the electron transport. Using the Transfer Hamiltonian approach to describe the tunneling current terms in ballistic regime, the rate equations can be used in order to obtain the non equilibrium distribution functions in each Qd. The effect of self-charge has been taken into account solving the Poisson equation with the appropriate boundary conditions for each Qd that involve the capacity coupling between the different parts of the system and the accumulated charge in each Qd. As expected, the calculation of the local potential inside each Qd is one of the most critical points, since the I-V curves depends on the position of the energy level. Due to the simplicity of the model, this can be easily extended to analyze arbitrary large arrays of Qds of interest in technological applications.  \smallskip

In order to simulate devices as realistic as possible, suitable expressions for the transmission coefficients, the energy level positions and the capacitive coupling have been used. These parameters can be described using basic material properties and geometrical representations of the system. Moreover, the hole currents have been taken into account, obtaining a complete description of the electron transport in the structure.  \smallskip

Finally, two prototypical structures have been simulated using realistic material parameters to describe an array of Si Qds embeeded in a $SiO_2$ insulator matrix. These structures compose the basic building blocks for future devices based in Qds ans demonstrate the practicability of the here presented approach. 
  
\section*{Acknowledgments}
S. Illera is supported by the FI programme of the Generalitat de Catalunya. A. Cirera acknowledges support from ICREA academia program. The authors thankfully acknowledge the computer resources, technical expertise and assistance provided by the Barcelona Supercomputing Center - Centro Nacional de Supercomputaci\'on. The research leading to these results has 
received funding from the European Community’s Seventh Framework Programme 
(FP7/2007-2013) under grant agreement n°: 245977.


\end{document}